\documentclass[12pt]{iopart}
\input psfig.sty
\begin{document}

\title[Fluctuation-dissipation relation for an Ising-Glauber-like model]
{A far-from-equilibrium fluctuation-dissipation relation
for an Ising-Glauber-like model}

\author{Christophe Chatelain\dag}
\address{\dag\ Laboratoire de Physique des Mat\'eriaux,
Universit\'e Henri Poincar\'e Nancy I,
BP~239, Boulevard des aiguillettes,
F-54506 Vand{\oe}uvre l\`es Nancy Cedex,
France}
\ead{chatelai@lpm.u-nancy.fr}

\begin{abstract}
We derive an exact expression of the response function to an infinitesimal
magnetic field for an Ising-Glauber-like model with arbitrary exchange
couplings. The result is expressed in terms of thermodynamic averages and does
not depend on the initial conditions or on the dimension of the space. The
response function is related to time-derivatives of a complicated correlation
function and so the expression is a generalisation of the equilibrium
fluctuation-dissipation theorem in the special case of this model.
Correspondence with the Ising-Glauber model is discussed.
A discrete-time version of the relation is implemented in Monte Carlo
simulations and then used to study the aging regime of the
ferromagnetic two-dimensional Ising-Glauber model quenched from the
paramagnetic phase to the ferromagnetic one. Our approach has the originality
to give direct access to the response function and the fluctuation-dissipation
ratio. 
\end{abstract}

\submitto{\JPA}
\maketitle

\def\build#1_#2^#3{\mathrel{
\mathop{\kern 0pt#1}\limits_{#2}^{#3}}}
\def\ket#1{\left| #1 \right\rangle}
\def\bra#1{\left\langle #1\right|}
\def\braket#1#2{\left\langle\vphantom{#1#2} #1\right.
\left| \vphantom{#1#2}#2\right\rangle}
\def\spring{\hskip 0pt minus 1fil}
\def\identite{{\rm 1}\hbox to 1 pt{\spring\rm l}}

\section{Introduction}
The knowledge about out-of-equilibrium processes is far from being as advanced
as for systems at thermodynamical equilibrium. In particular, the
fluctuation-dissipation theorem (FDT) which holds at equilibrium is known
to be violated out-of-equilibrium. This theorem states that at equilibrium
the response function $R^{\rm eq}(t-s)$ at time $t$ to an infinitesimal field
applied to the system at time $s<t$ is related to the time-derivative of
the two-time autocorrelation function $C^{\rm eq}(t-s)$:
\begin{equation}
  R^{\rm eq}(t-s)=\beta{\partial\over \partial s}C^{\rm eq}(t-s).
  \label{intro1}
\end{equation}
In the Ising case, the response function reads $R_{ji}(t,s)
={\delta\langle\sigma_j(t)\rangle\over\delta h_i(s)}$ and the
correlation function $C_{ji}(t,s)=\langle\sigma_j(t)\sigma_i(s)\rangle$.
Based on a mean-field study of spin-glasses,
Cugliandolo {\sl et al.}~\cite{Cugliandolo94} have conjectured
that for asymptotically large times the FDT can be generalised by adding a
multiplicative factor $X(t,s)$ which moreover depends on time only through
the correlation function:
\begin{equation}
  R(t,s)\build\sim_{t\sim s\gg 1}^{}
        \beta X(C(t,s)){\partial\over \partial s}C(t,s).
  \label{intro2}
\end{equation}
The quantity $\beta^{\rm eff}(t,s)=\beta X(C(t,s))$ is interpreted as an
effective inverse temperature. Exact results have been obtained for the
ferromagnetic Ising chain~\cite{Godreche00a, Lippiello00} that confirm this
conjecture. Unfortunately, the response function is rarely so easily
accessible for more complex systems.
Both numerically and experimentally, only the integrated response function
is usually measured by applying a finite magnetic field to the system during a
finite time. In the so-called TRM scheme, the magnetic field is applied between
the times $0$ and $s$ and the magnetisation is measured at time $t$. Assuming
the equation~(\ref{intro2}) valid for any times $t$ and $s$, one can relate the
integrated response function to the fluctuation-dissipation (FD) ratio:
\begin{equation}
  \chi(t,s)=\int_0^s R(t,u)du
        \sim\beta \int_{C(t,0)}^{C(t,s)} X(C)dC.
  \label{intro3}
\end{equation}
The FD ratio $X(t,s)$ can thus be obtained as the slope of the
integrated response function $\chi(t,s)$ when plotted versus the
correlation function $C(t,s)$.
This method has been applied to the numerical study of many systems:
2d and 3d-Ising ferromagnets~\cite{Barrat98}, 3d Edwards-Anderson
model~\cite{Franz95, Barrat98}, 3d and 4d-Gaussian Ising
spin-glasses~\cite{Marinari98}, 2d Ising ferromagnet with dipolar
interactions~\cite{Stariolo99}, Heisenberg anti-ferromagnet on the Kagome
lattice~\cite{Bekhechi03}~$\ldots$ The conjecture~(\ref{intro2}) has also
recently been checked experimentally for a spin-glass~\cite{Herisson02}.
More details may be found in the reviews~\cite{Cugliandolo02,
Crisanti02}. However, the integrated response function depends linearly on the
FD ratio only if the conjecture~(\ref{intro2}) holds, which has not been
demonstrated for any of the previously cited systems. We will see in the case
of the homogeneous Ising model that this approach may lead to misinterpretations
and erroneous values of $X(t,s)$. The generalisation of the equilibrium FDT
has recently become an increasingly popular issue. Let us mention two of them:
an approximate generalisation of the FDT to metastable systems~\cite{Baez03}
(limited to dynamics having a transition rate $W$ with only one negative
eigenvalue) that has been successfully compared to numerical data for the
2D-Ising model and a generalisation of the FDT for trap models~\cite{Ritort03}.

  In the present work, we study the dynamics of an Ising-Glauber-like model.
In the section 1, we describe the model and its dynamics which are
studied analytically in the section 2. The response function to an
infinitesimal magnetic field is exactly calculated far-from-equilibrium.
It turns out that the response function is no more related to a
time-derivative of the spin-spin correlation function but to time-derivatives
of a more complicated correlation function. The equilibrium limit is shown
to have the usual form.
In the section 3, a discrete-time version of this expression is
implemented in Monte Carlo simulations. Our approach presents several
advantages: (i) we can compute directly the response function and not only the
integrated response function, (ii) we obtain the response function to an
infinitesimal magnetic field so that we avoid non-linear effects due to the use
of a finite magnetic field, (iii) the FD ratio can be computed without
resorting to Cugliandolo conjecture~(\ref{intro2}) and (iv) we can calculate
the response function $R(t,s)$ and the FD ratio $X(t,s)$ for any time
$t$ and $s<t$ during one single Monte Carlo simulation.
We performed Monte Carlo simulations of the two-dimensional homogeneous Ising
model quenched at and below the critical temperature $T_c$. In both cases, the
expected scaling behaviour of the response function in the aging regime
is well reproduced by the numerical data. The value of the exponent $a$,
still controversial, is estimated and the FD ratio is computed.
Our estimate of $X_{\infty}$ at $T_c$ turns out to be compatible with
previous work and the scaling behaviour of $X(t,s)$ below $T_c$ is
well reproduced. In both cases, the FD ratio depend
on time not only through the correlation function.

\section{Our Ising-Glauber-like model}
\subsection{Useful relations on Markov processes}
We consider a classical Ising model whose degrees of freedom are $N$ scalar
variables $\sigma_i=\pm 1$ located at the nodes of a $d$-dimensional lattice.
Let us denote by $\wp(\{\sigma\},t)$ the probability to observe the system in
the state $\{\sigma\}$ at time $t$. We first define a discrete-time Markov
chain by the master equation
\begin{equation}
	\fl\wp(\{\sigma\},t+\Delta t)
	=(1-\Delta t)\wp(\{\sigma\},t)+\Delta t\sum_{\{\sigma'\}}
    	W(\{\sigma'\}\rightarrow \{\sigma\},t)\wp(\{\sigma'\},t)
  \label{eq1a}
\end{equation}
where $W(\{\sigma\}\rightarrow \{\sigma'\},t)$ is the transition rate per unit
time from the state $\{\sigma\}$ to the state $\{\sigma'\}$ at time $t$.
The condition $\sum_{\{\sigma'\}} W(\{\sigma\}\rightarrow\{\sigma'\},t)=1$
ensures the normation of the probability $\wp(\{\sigma\},t)$ at any time $t$.
The system is not forced to make a transition at each time step, i.e. the
transition rate may have non-zero diagonal elements
$W(\{\sigma\}\rightarrow\{\sigma\},t)$.
In the continuous-time limit $\Delta t\rightarrow 0$,
the master equation (\ref{eq1a}) goes to
\begin{equation}
  \left(1+{\partial\over\partial t}\right)\wp(\{\sigma\},t)
  =\sum_{\{\sigma'\}} W(\{\sigma'\}\rightarrow \{\sigma\},t)\wp(\{\sigma'\},t).
  \label{eq1}
\end{equation}
It is easily shown that the conditional probability,
$\wp(\{\sigma\},t|\{\sigma'\},s)$ with $s<t$, defined by the Bayes relation
\begin{equation}
        \wp(\{\sigma\},t)=\sum_{\{\sigma'\}}
        \wp(\{\sigma\},t|\{\sigma'\},s)\wp(\{\sigma'\},s)
        \label{eq2}
\end{equation}
satisfies the same master equation~(\ref{eq1a}) too:
\begin{eqnarray}
	\wp(\{\sigma\},t+\Delta t|\{\sigma'\},s)
	&=&(1-\Delta t)\wp(\{\sigma\},t|\{\sigma'\},s)	\nonumber\\
	\ms
	&+&\Delta t\sum_{\{\sigma''\}} W(\{\sigma''\}\rightarrow \{\sigma\},t)
        \wp(\{\sigma''\},t|\{\sigma'\},s)
	\label{eq3a}
\end{eqnarray}
or in the continuous-time limit
\begin{equation}
        \left(1+{\partial\over\partial t}\right)
        \wp(\{\sigma\},t|\{\sigma'\},s)
        =\sum_{\{\sigma''\}} W(\{\sigma''\}\rightarrow \{\sigma\},t)
   	\wp(\{\sigma''\},t|\{\sigma'\},s).
        \label{eq3}
\end{equation}
Moreover, one can work out a master equation for the time $s$. It reads
\begin{eqnarray}
	\wp(\{\sigma\},t|\{\sigma'\},s+\Delta t)
	&=&(1+\Delta t)\wp(\{\sigma\},t|\{\sigma'\},s)	\nonumber\\
	\ms
	&-&\Delta t\sum_{\{\sigma''\}} W(\{\sigma'\}\rightarrow \{\sigma''\},s)
	\wp(\{\sigma\},t|\{\sigma''\},s)
        \label{eq4a}
\end{eqnarray}
and in the continuous-time limit
\begin{equation}
        \left(1-{\partial\over\partial s}\right)
        \wp(\{\sigma\},t|\{\sigma'\},s)
        =\!\sum_{\{\sigma''\}} W(\{\sigma'\}\rightarrow \{\sigma''\},s)
	\wp(\{\sigma\},t|\{\sigma''\},s).
        \label{eq4}
\end{equation}
This last equation might be obtained simply by using for example the identity
${\partial\over\partial s}\wp(\{\sigma\},t)=0$.

When the transition rates do not depend on time, the conditional
probability $\wp(\{\sigma\},t|\{\sigma'\},s)$ is a function of $t-s$ only.
This can be shown easily by introducing the matrix notation
$\wp(\{\sigma\},t|\{\sigma'\},s)=\bra{\{\sigma\}}\hat\wp(t,s)\ket{\{\sigma'\}}$.
The master equation~(\ref{eq3}) reads then:
\begin{equation}
        {\partial\over\partial t}\hat\wp(t,s)
        =(\hat W-\identite)\hat\wp(t,s)
        \label{eq5}
\end{equation}
where $\bra{\{\sigma\}}\hat W\ket{\{\sigma'\}}=W(\{\sigma\}\rightarrow
\{\sigma'\})$. This equation admits the formal solution
\begin{eqnarray}
        \wp(\{\sigma\},t|\{\sigma'\},s)&=&\sum_{\{\sigma''\}}
        \bra{\{\sigma\}} e^{\int_s^t (\hat W-\identite)dt'}\ket{\{\sigma''\}}
        \wp(\{\sigma''\},s|\{\sigma'\},s)               \nonumber\\
        &=&\bra{\{\sigma\}} e^{(\hat W-\identite)(t-s)}\ket{\{\sigma'\}}
        \label{eq6}
\end{eqnarray}
where the initial condition $\wp(\{\sigma''\},s|\{\sigma'\},s)
=\delta_{\{\sigma''\},\{\sigma'\}}$ has been used. This dependence only
on $t-s$, even far-from-equilibrium, will be used latter in the calculation of
the response function.
        
\subsection{The model and its dynamics}
The Ising model is defined by its equilibrium probability distribution
$\wp_{\rm eq}(\{\sigma\})$ which reads with general exchange couplings:
\begin{equation}
        \wp_{\rm eq}(\{\sigma\})
        ={1\over {\cal Z}}e^{-\beta{\cal H}(\{\sigma\})}
        ={1\over {\cal Z}}e^{\beta\sum_{k,l<k} J_{kl}\sigma_k\sigma_l}.
        \label{eq7}
\end{equation}
where ferromagnetic couplings correspond to $J_{kl}>0$.
The condition of stationarity ${\partial\over\partial t}
\wp_{\rm eq}(\{\sigma\})=0$ leads according to the master
equation~(\ref{eq1}) to a constrain on the transition rates: 
\begin{equation}
  \sum_{\{\sigma'\}}\Big[
    \wp_{\rm eq}(\{\sigma'\})W(\{\sigma'\}\rightarrow \{\sigma\},t)
    -\wp_{\rm eq}(\{\sigma\})W(\{\sigma\}\rightarrow \{\sigma'\},t)
    \Big]=0.
  \label{eq8}
\end{equation}
The equation (\ref{eq8}) is satisfied when the detailed balance holds:
\begin{equation}
  \wp_{\rm eq}(\{\sigma'\})W(\{\sigma'\}\rightarrow \{\sigma\},t)
  =\wp_{\rm eq}(\{\sigma\})W(\{\sigma\}\rightarrow \{\sigma'\},t).
  \label{eq9}
\end{equation}
This last unnecessary but sufficient condition is fulfilled by
the heat-bath single-spin flip dynamics defined by the
following transition rates:
\begin{equation}
  W(\{\sigma\}\rightarrow \{\sigma'\},t)
  ={1\over N}\sum_{k=1}^N W_k(\{\sigma\}\rightarrow \{\sigma'\})
  \label{eq10}
\end{equation}
where the transition rate for a single spin-flip is
\begin{equation}
  W_k(\{\sigma\}\rightarrow \{\sigma'\})
  =\left[\prod_{l\ne k}\delta_{\sigma_l,\sigma'_l}\right]
  {e^{\beta\sum_{l\ne k} J_{kl}\sigma'_k\sigma'_l}\over
  \sum_{\sigma=\pm 1} e^{\beta\sum_{l\ne k} J_{kl}\sigma\sigma'_l}}.
  \label{eq10a}
\end{equation}
In this last expression, only the single-spin flip $\sigma_k\rightarrow
\sigma'_k$ is allowed. The product of Kronecker deltas ensures that all other
spins are not modified during the transition. After the transition, the spin
$\sigma_k$ takes the new value $\sigma'_k$ chosen according to the equilibrium
probability distribution $\wp_{\rm eq}(\{\sigma\})$. In the case of the Ising
chain, the transition rates (\ref{eq10}) are equivalent to Glauber's
ones~\cite{Glauber63}.
We will use a slightly different dynamics consisting in a sequential update of
spins. Let us choose a sequence of lattice sites $\{\kappa(t)\in\{1,\ldots N\}
,\forall t=n\Delta t,n\in {\rm I\!N}\}$ and let us define the transition
rates in discrete time as
\begin{equation}
	W(\{\sigma\}\rightarrow \{\sigma'\},t)
	=W_{\kappa(t)}(\{\sigma\}\rightarrow \{\sigma'\}).
  	\label{eq10b}
\end{equation}
In comparison to Glauber dynamics, only the spin-flip involving the spin
$\sigma_{\kappa(t)}$ is possible at time $t$. In the continuous-time limit,
the two dynamics are equivalent up to a rescaling of time
$t\rightarrow t/N$ (found for example in the definition of a Monte Carlo
step). Indeed, when iterating $N$ times the master equation (\ref{eq1a}),
one obtains
\begin{eqnarray}
	\fl\wp(\{\sigma\},t+N\Delta t)
	=&&(1-N\Delta t)\wp(\{\sigma\},t)	\nonumber\\
	\fl&&+\Delta t\sum_{\{\sigma'\}}
    	\wp(\{\sigma'\},t)\sum_{n=0}^{N-1} W_{\kappa(t+n\Delta t)}
	(\{\sigma'\}\rightarrow \{\sigma\})+\Or(\Delta t^2).
  	\label{eq10c}
\end{eqnarray}
and the Glauber dynamics is recovered if $\{\kappa(t+n\Delta t)\}_{n=0
\ldots N-1}$ is any circular permutation of the set of lattice sites
$\{1,\ldots N\}$. The equivalence of the two dynamics may not hold in the
thermodynamic limit $N\rightarrow +\infty$. The time-dependence of the
transition rates (\ref{eq10b}) breaks the time-translation invariance of the
conditional probabilities. However, the
effective transition rate in equation (\ref{eq10c}) is time-independent
and thus the time-translation invariance is restored in the continuous-time
limit if $\{\kappa(t)\}$ is periodic of period $N\Delta t$. Again, this
may be no more true in the thermodynamic limit. In the following, we will
assume that $\{\kappa(t)\}$ satisfies the two above-presented conditions, i.e.
being periodic of period $N\Delta t$ and that any $N$ consecutive values
are a circular permutation of $\{1,\ldots N\}$.

\section{Fluctuation-dissipation relation}
\subsection{Far-from-equilibrium fluctuation-dissipation relation}
A magnetic field $h_i$ is coupled to the spin $\sigma_i$ between the times
$s$ and $s+\Delta t$. During this interval of time, the transition rates
are changed to
\begin{equation}
  W_{k=\kappa(s)}^h(\{\sigma\}\rightarrow \{\sigma'\})
  =\left[\prod_{l\ne k}\delta_{\sigma_l,\sigma'_l}\right]
  {e^{\beta\left[\sum_{l\ne k} J_{kl}\sigma'_k\sigma'_l
      +h_i\sigma'_k\delta_{k,i}\right]}
  \over\sum_{\sigma=\pm 1} e^{\beta\left[
      \sum_{l\ne k} J_{kl}\sigma\sigma'_l+h_i\sigma\delta_{k,i}\right]}}
\label{eq11}
\end{equation}
in order to take into account the additional Zeeman term $\beta h_i\sigma_i$
in the Hamiltonian of the equilibrium probability distribution (\ref{eq7}).
The transition rates are all identical to the case $h_i=0$ apart from the
single-spin flip $W_i^h$ involving the spin $\sigma_i$.

Using the Bayes relation and
the discrete-time master equation (\ref{eq1a}), the average of the spin
$\sigma_j$ at time $t>s$ can be expanded under the following form:
\begin{eqnarray}
    \fl\langle\sigma_j(t)\rangle
    &=&\sum_{\{\sigma\}}\sigma_j\ \!\wp(\{\sigma\},t) 		\nonumber\\
    \fl&=&\sum_{\{\sigma\},\{\sigma'\}}\sigma_j\ \!
    \wp(\{\sigma\},t|\{\sigma'\},s+\Delta t)
	\wp(\{\sigma'\},s+\Delta t) 				\nonumber\\
    \fl&=&\sum_{\{\sigma\},\{\sigma'\}}\sigma_j\ \!
    \wp(\{\sigma\},t|\{\sigma'\},s+\Delta t)
    \Big[(1-\Delta t)\wp(\{\sigma'\},s)\Big.       		\nonumber\\
    \fl& &\hskip 110pt
    \Big.+\Delta t \sum_{\{\sigma''\}} W_{\kappa(s)}^h(\{\sigma''\}
	\rightarrow\{\sigma'\})\wp(\{\sigma''\},s)\Big]. 	
\label{eq12}
\end{eqnarray}
$W_i^h$ being the only quantity depending on the magnetic field in equation
(\ref{eq12}), only remains the second term when $\kappa(s)=i$ after derivating
with respect to the magnetic field . The derivative leads then to
\begin{eqnarray}
	\fl\left[{\partial \langle\sigma_j(t)\rangle\over\partial h_i}
  	\right]_{h_i\rightarrow 0}
	&=&\Delta t\ \!\delta_{\kappa(s),i}
	\sum_{\{\sigma\},\{\sigma'\},\atop\{\sigma''\}}\sigma_j\ \!
    	\wp(\{\sigma\},t|\{\sigma'\},s+\Delta t)
							\nonumber\\
	\fl& &\hskip 70pt\times\left[{\partial W_i^h\over\partial h_i}
	(\{\sigma''\}\rightarrow \{\sigma'\})\right]_{h_i\rightarrow 0}
	\wp(\{\sigma''\},s)
  \label{eq14}
\end{eqnarray}
This quantity is the magnetisation on site $j$ at time $t$ when an
infinitesimal magnetic field is applied to the site $i$ between $s$ and
$s+\Delta t$, i.e. an integrated response function that we will denote
$\chi_{ji}(t,[s;s+\Delta t])$. The derivative of the transition rate $W_i^h$
defined by equation (\ref{eq11}) is easily taken and reads
\begin{equation}
  \fl\left[{\partial W_i^h\over\partial h_i}
	(\{\sigma''\}\rightarrow \{\sigma'\})\right]_{h_i\rightarrow 0}\!\!\!\!
  =\beta W_i(\{\sigma''\}\rightarrow \{\sigma'\})\Big[
    \sigma'_i-\tanh\Big(\beta\sum_{k\ne i} J_{ik}\sigma_k'\Big)\Big].
  \label{eq15}
\end{equation}
It turns out to involve the transition rate of the zero-field
dynamics~(\ref{eq10a}). Due to this property, the integrated response function
can be expressed in terms of thermodynamic averages of the zero-field
dynamics. Inserting (\ref{eq15}) into (\ref{eq14}), the integrated response
function is rewritten as
\begin{eqnarray}
    \fl\chi_{ji}(t,[s;s+N\Delta t])
    &=&\beta\Delta t\ \!\delta_{\kappa(s),i}
	\sum_{\{\sigma\},\{\sigma'\},\atop\{\sigma''\}}
	\sigma_j\ \!\wp(\{\sigma\},t|\{\sigma'\},s+\Delta t)
								\nonumber\\
    \fl&\times&\Big[\sigma'_i
	-\tanh\Big(\beta\sum_{k\ne i} J_{ik}\sigma_k'\Big)\Big]
	W_i(\{\sigma''\}\rightarrow \{\sigma'\})
	\wp(\{\sigma''\},s)
  \label{eq16}
\end{eqnarray}
The summation over $\{\sigma''\}$ can be performed by using the discrete-time
master equation (\ref{eq1a}). One obtains
\begin{eqnarray}
    \fl\chi_{ji}(t,[s;s+N\Delta t])
    =\beta\ \!\delta_{\kappa(s),i}
	\sum_{\{\sigma\},\{\sigma'\}}&&\sigma_j\ \!
	\wp(\{\sigma\},t|\{\sigma'\},s+\Delta t)\Big[\sigma'_i
	-\tanh\Big(\beta\sum_{k\ne i} J_{ik}\sigma_k'\Big)\Big]
		\nonumber\\
    \fl&&\times
	\Big[\wp(\{\sigma'\},s+\Delta t)-(1-\Delta t)
		\wp(\{\sigma'\},s)\Big]
  \label{eq17}
\end{eqnarray}
Using a Taylor-expansion of $\wp(\{\sigma'\},s)$ in the vicinity
of $s+\Delta t$, equation (\ref{eq17}) can be rewritten to lowest order
in $\Delta t$ as
\begin{eqnarray}
    \fl\chi_{ji}(t,[s;s+N\Delta t])
    =\beta\Delta t\ \!\delta_{\kappa(s),i}
	\sum_{\{\sigma\},\{\sigma'\}}\sigma_j\ \!&&
    \wp(\{\sigma\},t|\{\sigma'\},s+\Delta t)\Big[\sigma'_i
	-\tanh\Big(\beta\sum_{k\ne i} J_{ik}\sigma_k'\Big)\Big]	
	\nonumber\\
    \fl&&\times\Big[\wp(\{\sigma'\},s+\Delta t)
	+{\partial\wp\over\partial s}(\{\sigma'\},s+\Delta t)\Big].
  \label{eq19}
\end{eqnarray}
The time-translation invariance of conditional probabilities
being restored in the continuous-time limit, they are function of $t-s$ only
and thus satisfy the property
\begin{equation}
	{\partial \wp\over\partial s}(\{\sigma\},t|\{\sigma'\},s)
	=-{\partial \wp\over\partial t}(\{\sigma\},t|\{\sigma'\},s).
\end{equation}
The term involving the time-derivative in equation (\ref{eq19}) can thus be
rewritten in the continuous-time limit as
\begin{eqnarray}
	\wp(\{\sigma\},t|\{\sigma'\},s){\partial\wp\over\partial s}
	(\{\sigma'\},s)=&&{\partial\over\partial s}\Big[
	\wp(\{\sigma\},t|\{\sigma'\},s)\wp(\{\sigma'\},s)\Big]	\nonumber\\
	&&\hskip 15pt
	\underbrace{-{\partial\wp\over\partial s}(\{\sigma\},t|
	\{\sigma'\},s)}_{=+{\partial\wp\over\partial t}
	(\{\sigma\},t|\{\sigma'\},s)}\wp(\{\sigma'\},s).	
  \label{eq20}
\end{eqnarray}
Moreover, the integrated response function $\chi_{ji}(t,[s;s+\Delta t])$
goes to the response function $R_{ji}(t,s)$ in the continuous-time
limit~:
\begin{equation}
	\chi_{ji}(t,[s;s+\Delta t])=\int_s^{s+\Delta t} R_{ji}(t,u)du
	=R_{ji}(t,s)\Delta t+\Or(\Delta t^2).
  \label{eq21}
\end{equation}
Combining equations (\ref{eq19}), (\ref{eq20}) and (\ref{eq21}),
the response function reads in the continuous-time limit
\begin{equation}
	R_{ji}(t,s)=\beta\left(1+{\partial\over\partial s}
                		+{\partial\over\partial t}\right)
        \langle\sigma_j(t)\left[\sigma_i(s)
        -\sigma_i^{\rm Weiss}(s)\right]\rangle\ \!\delta_{\kappa(s),i}
  \label{eq22}
\end{equation}
where $\sigma_i^{\rm Weiss}(s)=\tanh\Big(\beta\sum_{k\ne i}
J_{ik}\sigma_k(s)\Big)$ is the equilibrium value of the spin
$\sigma_i$ in the Weiss field created by all other spins at time $s$.
Relation (\ref{eq22}) generalises equation~(\ref{intro1}). The response
function $R_{ji}(t,s)$ turns out to be related to time-derivatives of the
correlation function of the spin $\sigma_j$ at time $t$ with the fluctuations
of the spin $\sigma_i$ at time $s$ around the equilibrium average 
$\sigma_i^{\rm Weiss}(s)$ of this spin in its Weiss field.
In this sense, this relation is still a
fluctuation-dissipation relation but valid far-from-equilibrium.
No assumption has been made on the dimension of the space or on
the set of exchange couplings $J_{kl}$ during the calculation. Moreover, it
applies for any initial conditions $\wp(\{\sigma\},0)$. The appearance of
the prefactor $1+{\partial\over\partial s}+{\partial\over\partial t}$ is not
related to the equilibrium probability distribution of the model but comes
from the Markovian properties of the dynamics. 
Generalised response functions are easily calculated along the same lines than
equation (\ref{eq22}). The second-order term for example reads
\begin{eqnarray}
\label{eq22b}
  \fl R^{(2)}_{kji}(t,s,r)
        &=&\left({\delta^2\langle\sigma_k(t)\rangle
            \over \delta h_j(s)\delta h_i(r)}\right)_{h\rightarrow 0}
        \quad\quad(t>s>r)						\\
    \fl&=&\beta\left(1+{\partial\over\partial s}+{\partial\over\partial t}\right)
	\left(1+{\partial\over\partial r}+{\partial\over\partial s}
	+{\partial\over\partial t}\right)\langle\sigma_k(t)
            \ \!\delta\sigma_j(s)\ \!\delta\sigma_i(r)\rangle
	\ \!\delta_{\kappa(r),j}\delta_{\kappa(s),i}		\nonumber
\end{eqnarray}
where $\delta \sigma_j(s)=\sigma_j(s)-\sigma_j^{\rm Weiss}(s)$.
Calculation of non-linear terms requires higher-order derivatives of the
transition rate as for example
\begin{eqnarray}
\label{eq22c}
  	R^{(2)}_{jii}(t,s,s)
        &=&\left({\delta^2\langle\sigma_j(t)\rangle
            \over \delta h_i^2(s)}\right)_{h\rightarrow 0}
        \quad\quad(t>s)\\
        &=&-2\beta^2\left(1+{\partial\over\partial s}
              +{\partial\over\partial t}\right)\langle\sigma_j(t)
            \sigma_i^{\rm Weiss}(s)\delta\sigma_i(s)\rangle\ \!\delta_{\kappa(s),i}.
        \nonumber
\end{eqnarray}
These relations are moreover easily extended to other models.
The relations (\ref{eq22}) to (\ref{eq22c}) hold for the
$\Or(n)$ or the $q$-state Potts for example where $\sigma_i(s)$ has to be
replaced by the local order parameter at time $s$ on the site $i$ and
$\sigma^{\rm Weiss}_i(s)$ by its average value in the Weiss field.
Since equations (\ref{eq22}) to (\ref{eq22c}) involve a constraint
on the sequence of spin-flips, their generalisation to the Ising-Glauber
model is not trivial. However, they will be of great interest for Monte Carlo
simulations.

\subsection{Equilibrium limit}
We will show in this section that the usual expression of the
FDT~(\ref{intro1}) is recovered in the equilibrium limit.
At equilibrium, the probability distribution $\wp_{\rm eq}(\{\sigma\})$ does
not depend on time. As a consequence, the integrated response function
can be written according to equation (\ref{eq19}) as
\begin{eqnarray}
    \fl\chi_{ji}^{\rm eq}(t,[s;s+\Delta t])
    =\beta\Delta t\ \!\delta_{\kappa(s),i}
	\sum_{\{\sigma\},\{\sigma'\}}&&\sigma_j\ \!
    	\wp(\{\sigma\},t|\{\sigma'\},s+\Delta t)		\nonumber\\
    &&\times\Big[\sigma'_i-\tanh\Big(\beta\sum_{k\ne i} J_{ik}
	\sigma_k'\Big)\Big]\wp_{\rm eq}(\{\sigma'\})		
  \label{eq23}
\end{eqnarray}
The hyperbolic tangent can be expressed in terms of the transition ratio
of the zero-field dynamics~(\ref{eq10a}):
\begin{eqnarray}
    \fl\tanh\Big(\beta\sum_{k\ne i}J_{ik}\sigma'_k\Big)
	\wp_{\rm eq}(\{\sigma'\})&=&\sum_{\{\sigma''\}} \sigma_i''
    	{\left[\prod_{k\ne i}\delta_{\sigma''_k,\sigma'_k}\right]
     	e^{\beta\sum_{k\ne i} J_{ik}\sigma'_i\sigma'_k}
    	\over \sum_{\sigma=\pm 1}
    	e^{\beta\sum_{k\ne i} J_{ik}\sigma\sigma'_k}}
    	{e^{\beta\sum_{k,l<k} J_{kl}\sigma''_k\sigma''_l}\over {\cal Z}}
	\nonumber\\
    \fl&=&\sum_{\{\sigma''\}}\sigma_i''W_i(\{\sigma''\}\rightarrow \{\sigma'\})
    	\wp_{\rm eq}(\{\sigma''\}).
  \label{eq24}
\end{eqnarray}
Inserting in equation (\ref{eq23}), the integrated response function reads
\begin{eqnarray}
  \label{eq25}
	\fl\chi_{ji}^{\rm eq}(t,[s;s+\Delta t])
	=&&\beta\Delta t\ \!\delta_{\kappa(s),i}
	\Big[\sum_{\{\sigma\},\{\sigma'\}}\sigma_j\ \!\wp(\{\sigma\},t|
	\{\sigma'\},s+\Delta t)\sigma_i'\wp_{\rm eq}(\{\sigma'\})\Big.	\\
	\fl&&\Big.-\sum_{\{\sigma\},\{\sigma'\},\atop\{\sigma''\}}
	\sigma_j\ \!\wp(\{\sigma\},t|\{\sigma'\},s+\Delta t)\sigma_i''
	W_i(\{\sigma''\}\rightarrow \{\sigma'\})
	\wp_{\rm eq}(\{\sigma''\})\Big].			\nonumber
\end{eqnarray}
The first term can be expressed as a thermodynamic average while in the second,
one needs to get rid first of the transition rate. The Kronecker delta
constrains the only possible spin-flip to involve site $i$ at time $s$.
As a consequence, $W_i(\{\sigma''\}\rightarrow \{\sigma'\})$ can be
replaced by $W_{\kappa(s)}(\{\sigma''\}\rightarrow \{\sigma'\})$ and
the master equation  (\ref{eq4a}) can be applied to equation (\ref{eq25}).
Moreover, one can show that
\begin{eqnarray}
  \label{eq26}
	\fl\wp(\{\sigma\},t|\{\sigma''\},s)
	=&&(1-\Delta t)\wp(\{\sigma\},t|\{\sigma''\},s+\Delta t)	\\
	\ms&&+\Delta t\sum_{\{\sigma'\}}
	W(\{\sigma''\}\rightarrow \{\sigma'\},s)
	\wp(\{\sigma\},t|\{\sigma'\},s+\Delta t)+\Or(\Delta t^2). \nonumber
\end{eqnarray}
This relation is obtained by first putting alone
$\wp(\{\sigma\},t|\{\sigma''\},s)$ in the left member
of the master equation (\ref{eq4a}) and then by iterating the relation to make
disappear $\wp(\{\sigma\},t|\{\sigma''\},s)$ in the right member.
Equation (\ref{eq26}) is then used to eliminate the transition rate from
equation (\ref{eq25})~:
\begin{eqnarray}
  \label{eq27}
	\fl\chi^{\rm eq}_{ji}(t,[s;s+\Delta t])
	=\beta\ \!\delta_{\kappa(s),i}\Big[\Delta t
	&&\sum_{\{\sigma\},\{\sigma'\}}\sigma_j\ \!\wp(\{\sigma\},t|
	\{\sigma'\},s+\Delta t)\sigma_i'\wp_{\rm eq}(\{\sigma'\})\Big.
							\nonumber\\
	\fl&&-\sum_{\{\sigma\},\{\sigma'\}}
	\sigma_j\wp(\{\sigma\},t|\{\sigma'\},s)
	\sigma_i'\wp_{\rm eq}(\{\sigma'\})			\\
	\fl&&+(1-\Delta t)\sum_{\{\sigma\},\{\sigma'\}}
	\sigma_j\wp(\{\sigma\},t|\{\sigma'\},s+\Delta t)
	\sigma_i'\wp_{\rm eq}(\{\sigma'\})\Big].		\nonumber
\end{eqnarray}
The two terms of order $\Delta t$ cancel and it remains only
\begin{equation}
	\chi^{\rm eq}_{ji}(t,[s;s+\Delta t])
	=\beta\ \!\delta_{\kappa(s),i}\langle\sigma_j(t)\left[
	\sigma_i(s+\Delta t)-\sigma_i(s)\right]\rangle_{\rm eq}
  \label{eq28}
\end{equation}
and in the continuous-time limit, one obtains equilibrium
fluctuation-dissipation:
\begin{equation}
	\fl R_{ji}^{\rm eq}(t,s)
	=\beta\ \!\delta_{\kappa(s),i}{\partial\over\partial s}
	\langle\sigma_j(t)\sigma_i(s)\rangle_{\rm eq}
	=\beta\ \!\delta_{\kappa(s),i}\langle\sigma_j(t)\Big[
	\sigma_i(s)-\sigma_i^{\rm Weiss}(s)\Big]\rangle_{\rm eq}
  \label{eq29}
\end{equation}
where the last member is simply equation (\ref{eq22}) at equilibrium.
One recovers the usual equilibrium fluctuation-dissipation relation
up to a Kronecker delta due the fact that the response function is
non-zero only for times at which a spin-flip involving the spin connected
to the magnetic field occurs.

\section{Monte Carlo simulations of the 2d-Ising model}
The discrete-time analogous of expression~(\ref{eq22}) of the response
function enables to study the aging displayed by the
Ising-Glauber model more accurately than in previous works that were based on
the numerical estimate of the integrated response function. In the first
part of this section, the algorithm is given. In the second part,
simulations of the Glauber dynamics of the two-dimensional Ising model during
a quench from the paramagnetic phase to the ferromagnetic one are presented.
The system is expected to display aging, associated with the existence of
growing domains corresponding to competing ferromagnetic states~\cite{Bray94}.
Reversible processes occur in the bulk of domains while domain wall
rearrangements are irreversible. We will distinguish between quenches
at the critical temperature $T_c$ and below.
In both cases, lattice sizes $128\times 128$, $256\times 256$ and
$362\times 362$ were simulated and the data averaged over $3000$, $10000$ and
$5000$ initial configurations respectively. For all data, error bars were
estimated as the standard deviation around the average value.

\subsection{Discrete response function}
During a Monte Carlo simulation, the time is a discrete variable and the
time step is set to $\Delta t=1$. Monte Carlo simulations implement indeed
the Markov process defined by the master equation (\ref{eq1a}) with the
choice $\Delta t=1$. Since simulations are always made on finite systems,
dynamics with sequential and parallel updates are equivalent in the large-time
limit up to a time-renormalisation corresponding to the definition of a Monte Carlo
Step (MCS). The response function can only be defined for continuous
time processes. However, the integrated response function during one spin-flip
$\sigma_i\rightarrow \sigma_i'$ is the best estimator for the response function
$R_{ij}(t,s)$ that we can define. Inserting $\Delta t=1$ into equation
(\ref{eq17}), the estimator of the response function is simply
\begin{equation}
   	\chi_{ji}(t,[s;s+1])
	=\beta\ \!\delta_{\kappa(s),i}\langle \sigma_j(t)\Big[\sigma_i(s+1)
	-\sigma_i^{\rm Weiss}(s+1)\Big]\rangle
  \label{eq30}
\end{equation}
where $\sigma_i^{\rm Weiss}(s)=\tanh\Big(\beta\sum_{k\ne i}
J_{ik}\sigma_k'(s)\Big)$. In the following, we will be interested only on
response functions of the form $R_{ii}(t,s)$. In order to reduce statistical
fluctuations, we have then estimated the response function $R(t,s)$ as the
average over all spin-flips during one MCS:
\begin{equation}
    R(t,s)={1\over N}\sum_{n=0}^{N-1} \chi_{\kappa(s+n)\kappa(s+n)}
	(t,[s+n;s+n+1])
  \label{eq31}
\end{equation}
The calculation of this quantity is quite simple. Let evolve the simulation
until time $s$. For each of the $N$ next spin-flips
$\sigma_i\rightarrow \sigma_i'$, store the quantity
$\sigma_i'-\sigma_i^{\rm Weiss}$. Note that $\sigma_i'$ may be equal to
$\sigma_i$ meaning that the system has not changed during this time step.
However, in strict application of equation (\ref{eq30}), one has nevertheless
to store $\sigma_i-\sigma_i^{\rm Weiss}$. After $N$ spin-flips, let the
system evolve again until time $t$. Calculate the response function for
each site $i$ by multiplying the quantity $\sigma_i'-\sigma_i^{\rm Weiss}$
stored by the new value of the spin $\sigma_i$ and add all these one-site
response functions. Repeat the simulation as many times as necessary and
average the results. The integrated response function can be easily
calculated by numerical integration of the response function.

The time-derivative of the correlation function ${\partial \over\partial s}
C_{ji}(t,s)$ at time $s$ can be estimated by $\langle
\sigma_j(t)\Big[\sigma_i(s+1)-\sigma_i(s)\Big]\rangle$.
Again, this quantity is averaged over all spin-flips during one MCS.
The FD ratio (\ref{intro2}) can be estimated as
\begin{equation}
	\fl X(t,s)={R(t,s)\over \beta{\partial \over\partial s}C(t,s)}
	={\sum_{n=0}^{N-1} \langle \sigma_{\kappa(s+n)}(t)\Big[
	\sigma_{\kappa(s+n)}(s+1)-\sigma_{\kappa(s+n)}^{\rm Weiss}(s+1)\Big]
	\rangle\over\beta\sum_{n=0}^{N-1} \langle \sigma_{\kappa(s+n)}(t)
	\Big[\sigma_{\kappa(s+n)}(s+1)-\sigma_{\kappa(s+n)}(s)\Big]\rangle}.
  \label{eq32}
\end{equation}

\subsection{Quench at the critical temperature}
During a quench at the critical temperature $T_c$, the asymptotic decay
of the correlation function has been conjectured to
be~\cite{Janssen89, Godreche02}
\begin{equation}
        C(t,s)\build\sim_{t,s\gg 1}^{}
        s^{-a_c}{\cal C}_c(t/s)
\label{mc8}
\end{equation}
where $a_c={2\beta\over \nu z_c}$ and ${\cal C}_c(x)$ is a scaling function
that asymptotically behaves as ${\cal C}_c(x)\build\sim_{x\gg 1}^{}
x^{-\lambda_c/z_c}$. $\lambda_c$ is the critical autocorrelation
exponent~\cite{Fisher88} and $z_c$ the dynamical exponent. Similarly,
the asymptotic behaviour of the response function is
\begin{equation}
        R(t,s)\build\sim_{t,s\gg 1}^{}
        s^{-1-a_c}{\cal R}_c(t/s)
\label{mc7}
\end{equation}
where the scaling function ${\cal R}_c(x)$ behaves asymptotically as
${\cal R}_c(x)\build\sim_{x\gg 1}^{} x^{-\lambda_c/z_c}$ too.
By integrating over $s$, one obtains a relation similar to~(\ref{mc7})
for the integrated response function that has been checked by large-scale
Monte Carlo simulations~\cite{Henkel01}.
However, the relation~(\ref{mc7}) is asymptotic so is not expected to hold for
the response function $R(t,s)$ with small values of $s$ that are the main
contribution to the integrated response function. As a consequence, it is
difficult to test the asymptotic behaviour of the response function in this
way. Our approach permits us to avoid these problems and to test directly
the relation~(\ref{mc7}).

\begin{center}
\begin{figure}
        \centerline{\psfig{figure=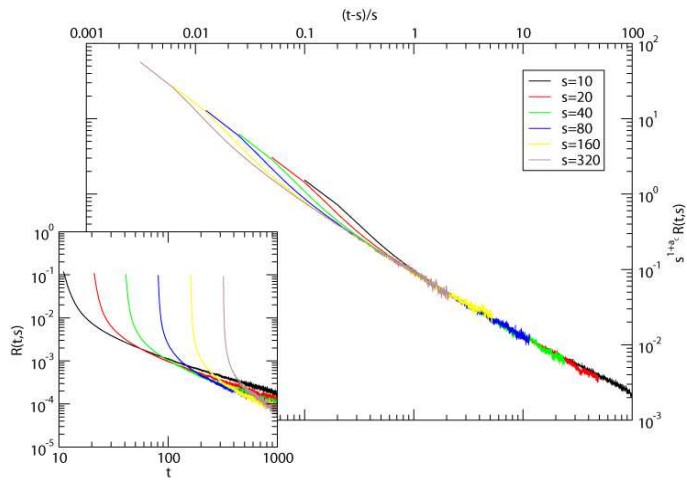,width=9cm}}
        \caption{Response function of the 2D-Ising model during a quench at
                the critical temperature (inset) and collapse of the scaling
                function for different values of $s$. The data were obtained
                with a lattice of size $362\times 362$ and averaged over
                5000 initial configurations.
                Each curve is surrounded by a clouds of dots corresponding to
                the lower and upper bounds of the error bars.
                The value $a_c=0.115$ was used~\cite{Godreche02}.
        }
        \label{fig1}
\end{figure}
\end{center}

The numerical data are plotted in figure~\ref{fig1}.
For the largest lattice size ($L=362$) and the smallest value of $s$ ($s=10$),
errors bars are at most $6\ \%$ of the value of the response function while
for the largest ($s=320$), they increase up to $12\ \%$. Indeed, in the
last case, the response function is of order of $1/N_{\rm config}$ and so can
not be sampled accurately. Nevertheless, a fairly good collapse of the data
is observed indicating that ${\cal R}_c(t/s)=s^{1+a_c}R(t,s)$ is indeed a
scaling function (actually we have used $(t-s)/s$ instead of $t/s$ but this
has no consequence on the asymptotic behaviour).

\begin{center}
\begin{figure}
	\centerline{\psfig{figure=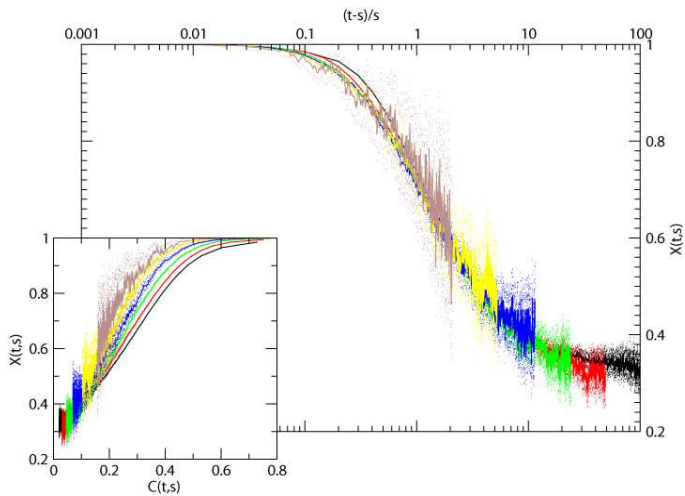,width=9cm}}
        \caption{FD ratio $X(t,s)$ versus $t$ for different values
          $s$ for the Ising-Glauber model quenched at the critical temperature
          $T_c$. The data were obtained for a lattice $362\times 362$ and
          averaged over 5000 initial configurations.
          Each curve is surrounded by a clouds of dots corresponding to
          the lower and upper bounds of the error bars.           
        }
        \label{fig3}
\end{figure}
\end{center}

We computed the FD ratio $X(t,s)$ using the estimator previously
derived and whose expression is given by equation~(\ref{eq32}).
The error bars are quite large. The numerical data are
plotted in figure~\ref{fig3} for the largest lattice size ($L=362$).
In contradistinction to Cugliandolo conjecture~(\ref{intro2}),
the inset of figure~\ref{fig3} shows that the FD ratio
does not depend on time only through the correlation function. However,
it seems that it may be the case in the limit $C(t,s)\rightarrow 0$.
On the other hand, it seems that the FD ratio depends on time only
through $t/s$ and reach a plateau for large enough values of $t$ that we may
estimate roughly to be $X_{\infty}\simeq 0.33(2)$. The same value is obtained
for $L=256$ and $L=362$ excluding any possibility of finite-size effects. The
limit $X_{\infty}$ has been conjectured to be universal~\cite{Godreche02} but
incompatible values have been given by different groups:
$X_{\infty}=0.26(1)$~\cite{Godreche00b} and
$X_{\infty}=0.340(5)$~\cite{Berthier03} by Monte Carlo simulations and
$X_{\infty}\simeq 0.35$~\cite{Gambassi02} for the $\Or(1)$-model in dimension
$d=4-\epsilon$. Our estimate is compatible with the last two ones.
The estimate $X_{\infty}=0.26(1)$ has probably been measured
for a too-short time $t$, far from the region where Cugliandolo
conjecture~(\ref{intro2}) and thus equation~(\ref{intro3}) hold. This puts
stress upon the danger of using equation~(\ref{intro3}) to compute the
FD ratio.

\subsection{Quench below the critical temperature}

\begin{center}
\begin{figure}
        \centerline{\psfig{figure=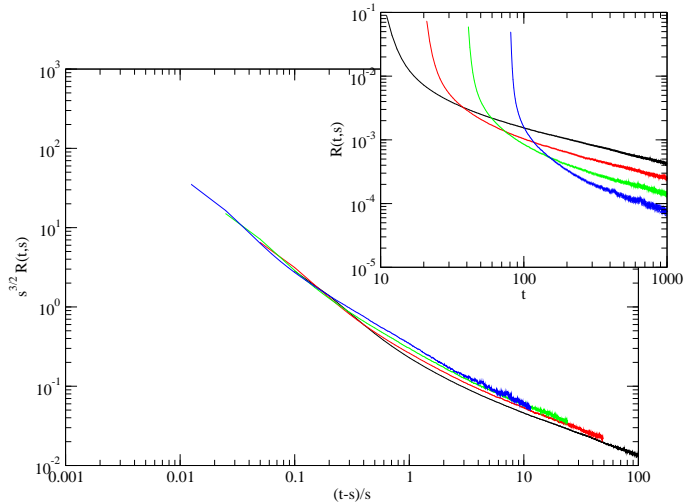,width=9cm}}
        \caption{Response function of the 2D-Ising model during a quench at
                the temperatures $T=J/0.6\simeq {3\over 4}T_c$ (inset) and
                collapse of the scaling function for different values of $s$.
                The data were obtained for a lattice $256\times 256$ and
                averaged over 10000 initial configurations.
                Each curve is surrounded by a clouds of dots corresponding to
                the lower and upper bounds of the error bars.           
        }
        \label{fig2}
\end{figure}
\end{center}

The same analysis can be done below $T_c$. In this regime, The
correlation function decays as~\cite{Janssen89, Godreche02}
\begin{equation}
        C(t,s)\build\sim_{t,s\gg 1}^{}
        M_{\rm eq}^2{\cal C}(t/s)
\label{mc9}
\end{equation}
where $M_{\rm eq}$ is the equilibrium magnetisation and ${\cal C}(x)$ a
scaling function that asymptotically behaves as
${\cal C}(x)\build\sim_{x\gg 1}^{} x^{-\lambda/z}$. The autocorrelation
exponent $\lambda$ and the dynamical exponent $z$ are expected to take
values which are different from those at $T_c$. The response function is
expected to scale as~\cite{Janssen89, Godreche02}
\begin{equation}
        R(t,s)\build\sim_{t,s\gg 1}^{}
        s^{-1-a}{\cal R}(t/s)
\label{mc10}
\end{equation}
where ${\cal R}(x)\build\sim_{x\gg 1}^{} x^{-\lambda/z}$. A controversy
exists concerning the value of $a$ that has been estimated to be either
$1/4$~\cite{Corberi02} or $1/2$~\cite{Henkel02,Paessens02}. Our numerical
data are presented in the figure~\ref{fig2}. We studied lattice sizes only up
to $L=256$ but calculations were made for two temperatures:
$J/0.6\simeq {3\over 4}T_c$ and $J/0.9\simeq T_c/2$. The error bars are much
smaller than in the critical case for small values of $s$. The relative error
is at most $2.6\ \%$ for $s=10$ at ${3\over 4}T_c$ but increase faster with
$s$: the relative error increases up to $11\ \%$ for $s=80$. As a
consequence, the study was limited to the values of $s$ ranging from $s=10$
to $s=80$. The response function displays the expected scaling behaviour
(\ref{mc10}) with $a=1/2$. However, the collapse is not perfect, especially for
the smallest values of $s$ but a very small variation of $a$ does not improve
it significantly. The value $a=1/4$ in particular improves the collapse for the
small values of $s$ only. The response function has probably strong corrections
to scaling. Note that corrections have already been taken into account for the
study of the scaling behaviour of the integrated response
function~\cite{Henkel01,Henkel03}. 

\begin{center}
\begin{figure}
        \centerline{\psfig{figure=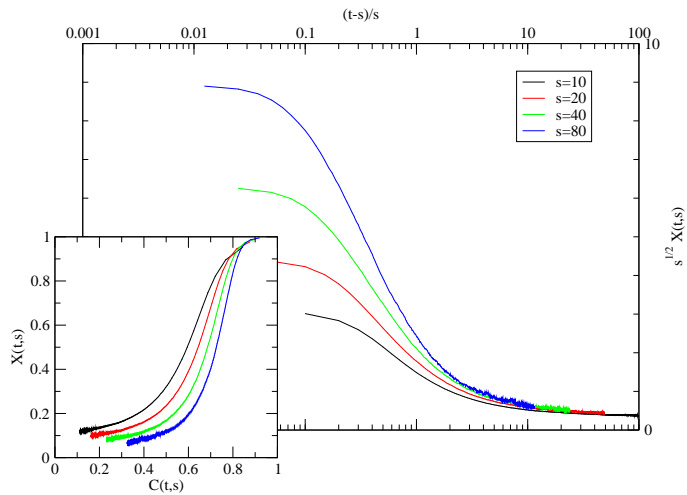,width=9cm}}
        \caption{FD ratio $X(t,s)$ versus $t$ for different values
          $s$ for the Ising-Glauber model quenched at the temperature
          $J/0.6k_B\simeq {3\over 4}T_c$.
          The data were obtained for a lattice $256\times 256$ and
          averaged over 10000 initial configurations.
          Each curve is surrounded by a clouds of dots corresponding to
          the lower and upper bounds of the error bars.           
        }
        \label{fig5}
\end{figure}
\end{center}

Combining the relations (\ref{mc9}) and (\ref{mc10}), the FD
ratio $X(t,s)=R(t,s)/\beta{\partial\over\partial s}C(t,s)$ is predicted to
vanish as $s^{-a}$ below $T_c$. Our numerical estimates for
$T=J/0.6\simeq {3\over 4}T_c$ are plotted in figure~\ref{fig5}. The
statistical errors decrease with the temperature so that the data are less
fluctuating than at $T_c$. As expected, the FD ratio is equal to
$1$ for small values of $t-s$, signalling that the main contribution to the
response function is due to equilibrium processes. On the other hand, it
vanishes in the limit $t\sim s\gg 1$ as $s^{-1/2}$. As shown in
figure~\ref{fig5}, the data for $s^{1/2}X(t,s)$ collapse for large values of
$t/s$. Moreover, figure~\ref{fig5} shows unambiguously that the FD
ratio does not depend on time only through the correlation function.
This makes the relation~(\ref{intro3}) invalid. The study of the
violation of the equilibrium FDT by the usual method relying on the
equation~(\ref{intro3}) would have led to erroneous values of the
FD ratio.

\section{Conclusion}
Using a formalism similar to Kubo's one in the quantum case, we derive an
exact expression of the response function of an Ising-Glauber-like model
far-from-equilibrium (equation~(\ref{eq22})). At least for finite systems,
the dynamics of our model is equivalent to the Glauber dynamics
up to a time-renormalisation $t\rightarrow t/N$. The derivation is possible
because the dynamics consists in a sequential update of the spins and the
transition rate under a magnetic field can be written as a product of the
transition rate without magnetic field and of a term depending only on the
final spin configuration. The response function turns
out to be related to time-derivatives of a correlation function involving the
fluctuations of the spin excited by the magnetic field around its equilibrium
average in its Weiss field. In this sense, the expression is a generalisation
of the equilibrium fluctuation-dissipation. Our expression
is quite general: no assumption has been made during its derivation on
the dimension of the space, the set of exchange couplings or the
initial conditions. Moreover, it can be easily extended to other classical
models. Generalised and non-linear response functions can be obtained analogously.
However, the continuous-time expression~(\ref{eq22}) may not hold in the
thermodynamic limit. Analytic results would be desirable. Unfortunately,
the response function calculated for the Ising-chain by Glauber itself in his
original paper~\cite{Glauber63} does not help because the magnetic field was
coupled differently to the system (by a multiplicative factor to make the
calculation feasible while we coupled the field by a modification of the
transition rate corresponding to the addition of the Zeeman interaction in the
equilibrium probability distribution). Generalisation to the Ising-Glauber model
is not trivial because equation $(\ref{eq29})$ sets a constrain on the sequence
of spin-flips. Nevertheless, It is tempting to imagine that like the
equilibrium FDT $(\ref{eq29})$, equations~(\ref{eq22}) to~(\ref{eq22c}) hold for
the Ising-Glauber model when suppressing this constrain on the sequence
of spin-flips.

The expression~(\ref{eq22}) of the response function is then implemented in
Monte Carlo simulations. Our approach gives access to the response function
and the FD ratio directly. In particular, the FD ratio can be obtained without
assuming the validity of the Cugliandolo conjecture~(\ref{intro2}).
We then study numerically the homogeneous two-dimensional Ising-Glauber model
quenched from the paramagnetic phase to the ferromagnetic one. Both the
response function and the FD ratio display the expected scaling
behaviour both at $T_c$ and below $T_c$. The values, still controversial, of
$a$ and $X_{\infty}$ are estimated to be equal to $1/2$ and $0.33(2)$
respectively, in agreement with some previous works.
The Cugliandolo conjecture~(\ref{intro2}) does not hold for this model apart
perhaps at $T_c$ in the limit of vanishing correlation functions. This would
explain discrepancies of previous estimates of $X_{\infty}$ relying on
Cugliandolo conjecture. The above-presented numerical procedure may be extended
to many different systems and would provided a unambiguous test of
Cugliandolo conjecture. We are currently studying the dynamics of
spin-glasses in this framework.
\medskip

\section*{Acknowledgements}
The laboratoire de Physique des Mat\'eriaux is Unit\'e Mixte de Recherche
CNRS number 7556. L'auteur remercie chaleureusement le groupe de physique
statistique du laboratoire de Physique des Mat\'eriaux de Nancy et tout
sp\'ecialement Dragi Karevski et Lo{\"\i}c Turban pour une relecture attentive
du manuscript.

\end{document}